\documentclass[10pt]{article}
\usepackage[T2A]{fontenc}
\usepackage[utf8]{inputenc}
\usepackage[english]{babel}
\usepackage{amsmath}
\usepackage{amssymb}
\usepackage{amsthm}
\usepackage{graphicx}
\usepackage{epstopdf}
\usepackage[affil-it]{authblk}

\usepackage{hyperref}
\hypersetup{
colorlinks=true,
linkcolor=red,
citecolor=blue,
urlcolor=blue
}

\usepackage[a4paper, total={7in, 9.6in}]{geometry}

\usepackage{multicol}
\usepackage{wrapfig}

\usepackage{titlesec}
\titleformat{\section}
{\normalfont\large\bfseries}
{\thesection.}{0.5em}{}
\titlespacing{\section}{0pc}{1pc}{0.5pc}

\def\C{\mathbb{C}}
\newcommand{\pd}[2]{\dfrac{\partial #1}{\partial #2}}

\newcommand{\bra}[1]{\langle #1 |}
\newcommand{\ket}[1]{| #1 \rangle}

\def\L{\mathcal{L}}     %superoperator
\def\Tr{{\rm Tr}}       %trace symbol Tr
\def\H{\mathcal{H}}     %Hilbert space
\title{\textbf{Incoherent GRAPE (inGRAPE) for optimization of \\quantum systems with environmentally assisted control}}

\date{}

\author[1,2,*]{Vadim N. Petruhanov}
\author[1,2,**]{Alexander N. Pechen}

\affil[1]{\it \normalsize Department of Mathematical Methods for Quantum Technologies,\par
Steklov Mathematical Institute of Russian Academy of Sciences,\par
8~Gubkina str., Moscow, 119991, Russia, }
\affil[2]{\it
University of Science and Technology MISIS,\par
6~Leninskiy prospekt, Moscow, 119991, Russia;}
\affil[*]{vadim.petrukhanov@gmail.com, \href{http://www.mathnet.ru/eng/person176798}{mathnet.ru/eng/person176798}}
\affil[**]{apechen@gmail.com, \href{http://www.mathnet.ru/eng/person17991}{mathnet.ru/eng/person17991}}

\begin{document}

\maketitle

\begin{abstract} 
In this work, we review several results on development and application of incoherent version of GRAPE (Gradient Ascent Pulse Engineering) approach, inGRAPE, to optimization for open quantum systems driven by both coherent and incoherent controls.\footnote{This work is based on the talk presented at the 15th International Conference “Micro- and Nanoelectronics --- 2023”  October 2--6, 2023, Zvenigorod, Russia. Since it is a brief overview of several recent results, we do not provide detailed references to various related works of many researchers. A more detailed overview of other related works can be found in the cited references.} In the incoherent control approach, the environment serves as a control together with coherent field, and decoherence rates become generally time-dependent. For a qubit, explicit analytic expressions for evolution of the density matrix were obtained by solving a cubic equation via Cardano method. We discuss applications of incoherent GRAPE method to high fidelity gate generation for open one- and two-qubit systems and surprising properties of the underlying control landscapes, forming two groups --- smooth single peak landscapes for Hadamard, C-NOT and C-Z gates, and more complicated with two peaks for T (or $\pi/8$) gate. For a qutrit, a formulation of the environment--assisted incoherent control with time-dependent decoherence rates is provided.
\end{abstract}

\bigskip


\bigskip

\begin{multicols}{2}

\section{Introduction}

Modern quantum technologies rely on many tools, one of which is quantum control that studies methods for manipulation of individual quantum systems~\cite{KochEPJQuantumTechnol2022}. Quantum control is also used for laser chemistry and optical control of molecular processes~\cite{RiceZhaoBook,TannorBook,Shapiro_Brumer_2011}. One of the important directions in quantum control is related to control of open quantum systems. Its importance is determined by the fact that in experimental circumstances quantum systems often can not be fully isolated from the environment. Moreover, in some cases the environment can be applied as a useful resource of control via, e.g., incoherent control~\cite{Pechen_Rabitz_2006, PechenPRA2011}.

Many quantum control problems are formulated as optimization of some objective functionals. In practice, various numeric optimization methods are  used for finding optimal shape of the control, such as genetic algorithms~\cite{Judson1992}, BFGS~\cite{EitanPRA2011}, Krotov~\cite{BaturinaMorzhinAiT2011}, CRAB~\cite{CanevaPRA2011}, and others. One of them is the GRadient Ascent Pulse Engineering (GRAPE) developed originally for design of NMR pulse sequences~\cite{khaneja_optimal_2005},  and later applied to various quantum control problems, e.g.,~\cite{deFouquieres2011, PechenTannor2012}.

A key point of the GRAPE approach is to derive explicit expressions for the gradient of the control objective. In this note, we discuss some work on the adaptation of the GRAPE method for open quantum systems driven by coherent and incoherent controls~\cite{PetruhanovPechenJPA2023}. We obtained analytic expressions for gradient and Hessian of Mayer-type objective functionals with respect to piecewise constant controls and $L_2$-controls for general $N$-level quantum systems. For one-qubit system, we managed to diagonalize matrix exponentials by solving the third order characteristic equation. We applied this developed incoherent GRAPE (inGRAPE) approach to the state transfer and the gate generation problems for the one-qubit system and for the two-qubit system~\cite{PetruhanovPhotonics2023,PetruhanovPechen_2023-2,Petruhanov2022}. High efficiency of the method allows to make a large amount (L = 1000) of launches and statistically analyze objective functionals (quantum control landscapes) for considered problems, e.g., on presence of local but not global optima (traps). While the method was applied before to one -and two-qubit systems, here we also provide a formulation of incoherent control for a qutrit (three-level quantum system) with one forbidden transition.

\section{Environmentally assisted quantum control and objectives}
Consider, generally, a (GKSL) master equation with coherent and incoherent control which describes evolution of an open quantum system~\cite{Pechen_Rabitz_2006}:
\begin{equation}\label{Eq:ME}
\pd{\rho}{t} = \mathcal{L}(\rho) = -i [H_0 + \sum_{k=1}^{K} u_k(t) V_k, \rho] + \L_{n(t)}(\rho),
\end{equation}
where $\rho$ is the system density matrix, $\rho \in \C^{N\times N}$, $\rho^\dagger = \rho \ge 0$, $\mathrm{Tr}\rho = 1$; $H_0$ is the free Hamiltonian; $V_k = V_k^\dagger$ is the interaction Hamiltonian; $u(t) = \{u_k(t)\}_{k = 1}^K$ is the coherent control, $u_k(t)$ are real-valued (physically, e.g., laser pulses); $n(t) = \{n_{ij}(t)\}$, $1 \le i < j \le \mathrm{dim\H}$, is the incoherent control, $n_{ij}(t) \geqslant 0$ are non-negative real-valued functions (physically {\em spectral density}), and $\mathcal{L}_{n}(\cdot): \C^{N\times N} \to \C^{N\times N}$ is the dissipative superoperator, which can have different forms corresponding to the following limits: {\em weak coupling limit}~\cite{Davies1976, Accardi_Volovich_Lu} and {\em low density limit}.

The considered control goal is to optimize Mayer-type  objective functionals:
\begin{equation}\label{Objective}
F[u,n] = J(\rho^{u,n}(T)) \to \inf_{u,n}.
\end{equation}
For example, the following objectives are considered in quantum control:
\begin{itemize}
\item observable $\mathcal{O}$ mean value:
$F_\mathcal{O}[u,n] = \Tr\,\mathcal O\rho^{u,n}(T);$
\item state-to-state transfer $\rho_0 \to \rho_{\rm target}$: $F_{\rho_{\rm target}}[u,n] = \|\rho^{u,n}(T) - \rho_{\rm target}\|^2 \to \inf_{u,n};$

\item unitary gate $U$ generation in an open system:\\ 
-- functional on states ${\rho_0^{(j)}}$:\\
$F_{U,K}\bigl(u, n; \rho_0^{(1)}, \dotsc ,\rho_0^{(K)}\bigr)= \\ \frac{1}{K}\sum_{j = 1}^K \left\| \Phi(T, u, n) \rho_0^{(j)} - U\rho_0^{(j)}U^\dagger \right\|^2\to \inf\limits_{u,n};$\\
-- functional on quantum channels:\\
$F_U(u, n) = \|\Phi(t, u, n) - U \cdot U^\dagger\|^2 \to \inf\limits_{u,n};$\\
where $\Phi(t,u,n)$ is the evolution operator.
\end{itemize}

\section{GRAPE for coherent and incoherent controls and applications to one- and two-qubit systems} 

In~\cite{PetruhanovPechenJPA2023}, we developed incoherent version of GRAPE approach for optimizing objective functionals for $N$-level open quantum systems driven by both coherent and incoherent controls.  To develop this incoherent version of GRAPE, we computed gradient of various objectives for general $N$-level open quantum systems in the class of piecewise constant controls. The case of a single qubit was considered in details and solved analytically. For this case, evolution equation using Bloch vector was formulated. Then an explicit analytical expression for the evolution, and hence for gradient of the objectives, was obtained using diagonalization of some $3 \times 3$ matrix determining  the system’s dynamics in the Bloch ball. This diagonalization was obtained by solving a certain cubic equation via Cardano method. The efficiency of the algorithm was demonstrated through numerical simulations for the state-to-state transition problem and its complexity was estimated. Finally, robustness of the obtained controls was studied. 

In~\cite{PetruhanovPhotonics2023}, this inGRAPE approach was applied to generation of single-qubit gates for a two-level open quantum system driven by coherent control and incoherent environment. The control problem was formulated as minimization of the objective
functional defined as the sum of Hilbert-Schmidt norms between four fixed basis states evolved under
the GKSL master equation with controls, and the same four states evolved under the ideal gate
transformation. An exact expression for the gradient of the objective functional with respect to
piecewise constant controls was obtained. Then a subsequent optimization was performed using the inGRAPE approach with an adaptive step size. As a result, optimal trajectories in the Bloch ball for various initial states were computed. In addition, a relation of the quantum gate generation problem with optimization on complex Stiefel manifolds was discussed. 

In~\cite{PetruhanovPechen_2023-2}, the efficiency of the developed inGRAPE approach for single-qubit gate generation was estimated using the numerical analysis of the corresponding quantum control landscapes and considering various objectives. For the analysis of the efficiency, we studied distribution of optimized by inGRAPE infidelity values and the corresponding controls. For the 
Hadamard gate, the distribution was found to have a simple form with one peak so the corresponding landscape is smooth. For T gate, a completely different situation was found with two peaks corresponding to two optimized infidelity values, one is smaller and one is larger. Hence for this gate inGRAPE can converge, depending on the initial control, to two different infidelity values that indicates the possibility in this case to have not only global minimum, but also a local  minimum in the underlying quantum control landscape. Similarly to the case of Hadamard gate, smooth single peak control landscapes were also found using inGRAPE for two-qubit C-NOT and C-Z gates~\cite{Pechen_etal}.

Applications to state transfer for a two-qubit system were studied in~\cite{Petruhanov2022}. 

\end{multicols}

\begin{figure}
    \centering
    \includegraphics[width = 0.5\linewidth]{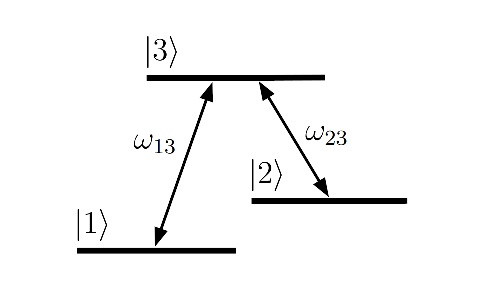}
    \caption{Energy levels of a qutrit with forbidden transition between states $|1\rangle$ and $|2\rangle$.}
    \label{fig:Fig1}
\end{figure}

\begin{multicols*}{2}

\section{Qutrit}

As the next step the method can be applied to control of such systems as qutrits ($N=3$). Qutrits together with qubits can be used for various tasks of information processing, e.g.~\cite{Popov_Kiktenko_Fedorov_Manko_2016}. 
For example, consider a special class of qutrits with three states $|1\rangle$, $|2\rangle$, $|3\rangle$ and with forbidden direct transition between states $|1\rangle$ and $|2\rangle$ (Fig.~\ref{fig:Fig1}). 
% \begin{figure*}
%     \includegraphics[width = 0.5\textwidth]{Fig1.eps}
%     \caption{Energy levels of a qutrit with forbidden transition between states $|1\rangle$ and $|2\rangle$.}
%     \label{fig:Fig1}
% \end{figure*}
Then in the non-degenerate case, free Hamiltonian of the qutrit and Hamiltonian for its interaction with coherent control have the form 
\begin{equation}
H_0= 
\begin{pmatrix}
E_1 & 0 & 0\\
0&  E_2 & 0\\
0& 0 & E_3
\end{pmatrix},\quad 
V=\begin{pmatrix}
0 & 0 & v_{13}\\
0 & 0 & v_{23}\\
v^*_{13} & v^*_{23} & 0
\end{pmatrix},
\end{equation}
\noindent where $E_1, E_2, E_3$ are (all different) energies of the states $|1\rangle$, $|2\rangle$, $|3\rangle$. There are two non-trivial transition frequencies $\omega_{13}=E_3-E_1$ and $\omega_{23}=E_3-E_2$. Hence incoherent control for this system has only two independent components $n_1(t)=n_{\omega_{13}}(t)\ge 0$ and $n_2(t)=n_{\omega_{23}}(t)\ge 0$ corresponding to density of particles of the surrounding environment (e.g., photons) with frequencies $\omega_{13}$ and $\omega_{23}$, respectively. 
Thus the dissipator describing evolution of this system under influence of incoherent control in the weak coupling limit has the form
\begin{gather*}
{\cal L}_{n(t)}(\rho)=n_1(t)A_1\Big(\bra{3}\rho\ket{3}\ket{1}\bra{1} + \bra{1}\rho\ket{1}\ket{3}\bra{3} \\- \{\ket{1}\bra{1} + \ket{3}\bra{3}, \rho\}\Big) + n_2(t)A_2\Big(\bra{3}\rho\ket{3}\ket{2}\bra{2}\\ + \bra{2}\rho\ket{2}\ket{3}\bra{3} - \{\ket{2}\bra{2} + \ket{3}\bra{3}, \rho\}\Big) \\+ A_1\Big(\bra{3}\rho\ket{3}\ket{1}\bra{1} - \{\ket{3}\bra{3}, \rho\}\Big)\\+A_2\Big(\bra{3}\rho\ket{3}\ket{2}\bra{2} - \{\ket{3}\bra{3}, \rho\}\Big),
\end{gather*}
where $A_1,A_2>0$ are the Einstein coefficients for spontaneous emission ($A_1$ is for transitions between levels $|3\rangle$ and $|1\rangle$, $A_2$ is for transitions between levels $|3\rangle$ and $|2\rangle$) and $\{\cdot,\cdot\}$ denotes anticommutator. 
Piecewise constant controls have the form:
\begin{equation}
u(t) = \displaystyle \sum_{k=1}^M u_k \chi_{[t_{k-1}, t_k)}(t),\quad  n_i(t) = \displaystyle \sum_{k=1}^M n^i_k \chi_{[t_{k-1}, t_k)}(t),
\end{equation}
where $i=1,2;$ $t \in [0,T]$; $\chi_{[t_{k-1},t_k)}$ is a characteristic function of the interval $[t_{k-1},t_k)$. The constraints $n_i\geq 0$ imply existence of the boundary of the control space which can be crossed during the optimization process. To avoid such crossing we introduce the change of the variable $n_i = w_i^2$, where $w_i\in\mathbb R$. Without such change of the variables, control of qutrit with coherent and incoherent drive was considered in~\cite{MorzhinQutrit} using GPM and Krotov-type methods. However, for inGRAPE such change of the variable via introduction of the new controls $w_i$ is a necessary crucial step.

Using these Hamiltonians, dissipator, and controls, one can compute the resulting evolution of the system density matrix using the master equation~(\ref{Eq:ME}) that in turn can be used to define control objectives of the form~\ref{Objective}. The detailed analysis of this model is a prospective for a future work.

\section{Conclusions}
We review some results on development and applications of incoherent GRAPE (inGRAPE) approach to optimization in open quantum systems driven by coherent and incoherent controls. The controls are considered in the functional class of piecewise constant controls. Thereby in the series of works we studied the dynamics and optimization, including derivation of gradients and Hessians, of various objectives for open one-qubit and two-qubit quantum systems with piecewise constant coherent and incoherent controls. For the one-qubit case, we diagonalized the matrix exponentials in the derived expressions by finding eigenvalues and eigenvectors of the $3\times 3$ matrix determining the dynamics in the Bloch parameterization. Derived expressions were applied to a numerical optimization in the context of the state transfer problem for one- and two-qubit systems. Beyond one- and two-qubit systems, based on the general incoherent control approach in this work we derive an explicit formulation of the environmentally--assisted incoherent control with time-dependent decoherence rates for a special class of qutrits with forbidden transition between ground and intermediate levels.

\section*{Acknowledgments}

Section 4 of this work was performed in Steklov Mathematical Institute of Russian Academy of Sciences within the Russian Science Foundation grant no. 22-11-00330, \href{https://rscf.ru/en/project/22-11-00330/}{https://rscf.ru/en/project/22-11-00330/}.

\bibliographystyle{unsrturl}
\bibliography{bibliography.bib}

\end{multicols*}

\end{document}